\documentclass[%
reprint,
showpacs,
nofootinbib,
aps,
pre,
floatfix,
]{revtex4-1}

\usepackage{graphicx}
\usepackage{dcolumn}
\usepackage{bm}

\usepackage{amssymb,amsmath,amsfonts} \usepackage{bm} 
\usepackage{graphicx, color} 

\definecolor{Myorange}{cmyk}{0,0.42,1,0}

\begin{document}
\title{In defence of the simple: Euclidean distance for comparing complex networks}

\author{Johann H. Mart\'inez}\email[]{johemart@gmail.com}\author{Mario Chavez}
\affiliation{INSERM-U1127, CNRS UMR7225, Sorbonne Universit\'e, ICM-H\^opital Piti\'e Salp\^etri\`ere-. Paris, France}
\date{\today}
\begin{abstract}
To improve our understanding of  connected systems, different  tools derived from statistics, signal processing, information theory and statistical physics have been developed in the last decade. Here, we will focus on the graph comparison problem. Although different estimates exist to quantify how different two networks are, an appropriate metric has not been proposed. Within this framework we compare the performances of different networks distances (a topological descriptor and a kernel-based approach) with the simple Euclidean metric. We define the performance of metrics as the efficiency of distinguish two network's groups and the computing time. We evaluate these frameworks on synthetic and real-world networks (functional connectomes from Alzheimer patients and healthy subjects), and we show that the Euclidean distance is the one that efficiently captures networks differences in comparison to other proposals. We conclude that the operational use of  complicated methods can be justified only by showing that they out-perform well-understood traditional statistics, such as Euclidean metrics.
\end{abstract}

\maketitle
Despite the success of complex networks modeling and analysis, some methodological challenges are still to be tackled to describe and compare different interconnected systems. Identifying and quantifying dissimilarities among networks is a challenging problem of practical importance in many fields of science. Given two graphs $\{G,\ G^{'}\}$, we aim at finding a real-valued function $f$ that maps $G\times G^{'}\to\mathbb{R} \ \forall\ \{G,\ G^{'}\}$. Functions $f(G,G^{'})$ that quantify the (dis)similarity between two networks  have been been studied in several areas such as chemistry, protein structures, social networks up to neuroscience, among others \cite{Borgwardt2005, Deshpande2005, Ralaivola2005}. Without an $f$ uniqueness, different approaches have been proposed including  isomorphisms, distances based on divergences, spectral parameters, kernels, or different combinations of the previous~\cite{Donnat2018, Wegner2018, Hammond2013, Schieber2017, Bai2015}. 

In this work, we consider three classes of  the function $f$:  the first class, who is the large bunch in the literature, quantifies local changes via structural differences. These metrics may range from the simplest Euclidean distance~\cite{Higham2002, Golub1996, Real1996} to more elaborated algorithms that assign costs of different operations to map nodes/edges of $G$ to their $G^{'}$ counterparts \cite{Sanfeliu1983}. Another distance class considers topological descriptors that map each graph into a feature vector (e.g. degree distribution, nodes centrality, etc.). These vectors are compared with any multivariate statistical distance to compute the graph dissimilarity \cite{Basseville1999, Runber1998, Schieber2017}. We notice that considering one type of feature may imply to lose topological information from others parameters, and the price of complet caracterisation may be paid with more runtime. The last class considered here includes kernel-based approaches that compare global substructures (i.e. walks, paths, etc). These methods capture global information of networks (e.g. the  graph Laplacian) considered in a metric space, where a defined inner product directly estimates its dissimilarity. Kernel methods, however, often integrates over local neighborhoods, which renders these approaches less sensitive to small or local perturbations~\cite{Donnat2018}. 

\begin{figure}[ht]
\centering
\resizebox{0.9\columnwidth}{!}{\includegraphics{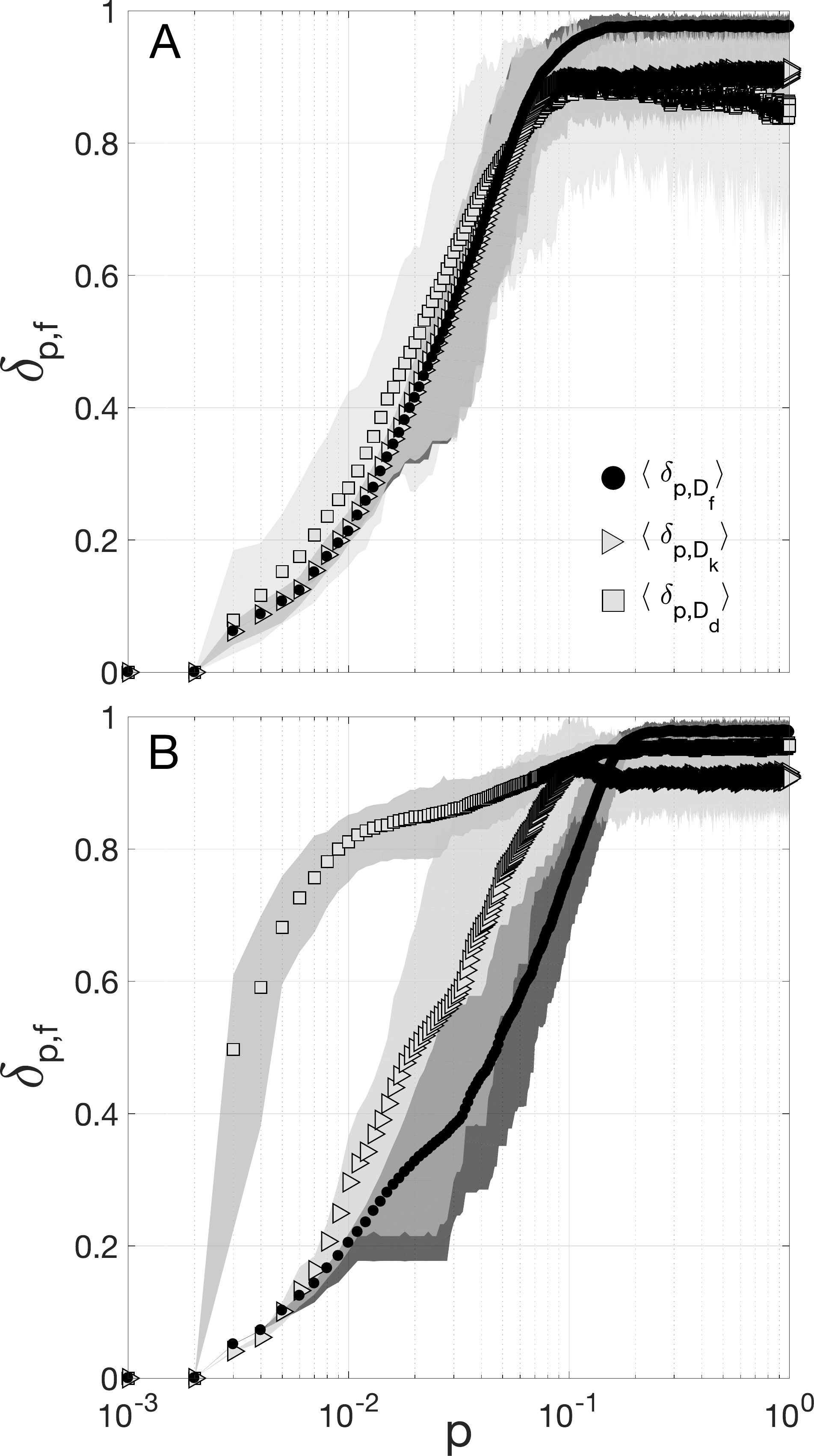}}
\caption{Network-distances as a function of the rewiring probability $p$. For visualisation purposes, each profile was normalized by dividing $\delta_{p,f}$ by the maximum value obtained over the whole range of perturbations. Black dots are the averages over $100$ realizations, and shaded areas indicate theirs 5$^{th}$ and 95$^{th}$ percentiles. \textbf{A}:  for the BA networks and \textbf{B}: for the SW model. In both models, $\langle\delta_{p,f}\rangle$ values were estimated for the same rewiring probability.}\label{fig:01}
\end{figure}

In this work  we show than the use of a simple Euclidean metric may provides good performances to asses graph differences, when compared to other more complicated functions. We propose a framework for measuring the performance of functions $f$'s applied on undirected-binary graphs of equal sizes. We define the $f$'s performance in terms of ``discriminability'' and ``runtime''. The former is the capability of $f$ for discriminating two sets of network associated to two different groups. The latter is simply the computing time. 

In what follows, we compare the performance of the standard Euclidean distance ($D_f$), the dissimilarity measure ($D_d$) defined in Ref~\cite{Schieber2017}, and the graph diffusion kernel distance ($D_k$)~\cite{Hammond2013}, from each of the classes mentioned above. For these algorithms, we evaluate the discriminability and runtime in different synthetic and real-world brain networks. We show that the Euclidean distance substantially outperforms other methods to capture differences between networks of the same size.

\emph{Euclidean distance.--} Assuming that $\{A_1, A_2\}$ are the adjacency matrix representations of graphs $\{G_1, G_2\}$, we have the Euclidean distance defined by:
\begin{equation}
\label{dE}
D_f = \|A_1-A_2\|_F
\end{equation}
where $\|\cdot\|_F$ denotes the Frobenius norm.

\emph{Network structural dissimilarity.--} This dissimilarity measure captures several topological descriptors~\cite{Schieber2017}: network distance distributions $\mu_{\{A_1,A_2\}}$, node-distance distribution functions $NND_{\{A_1,A_2\}}$ (local connectivity of each node), $\alpha$-centrality distributions $P_{\alpha \{A_1,A_2\}}$, the equivalent for their graph complements $P_{\alpha \{A_1^c,A_2^c\}}$ and several tuning parameters $\{\alpha, w_1,w_2,w_3\}$. The network distance is obtained via the Jensen-Shannon divergence $\Gamma$ between different feature vectors.

\begin{eqnarray}
D_d = &w_1 \sqrt{\frac{\Gamma(\mu_{A_1}, \mu_{A_2})}{log 2}}+ w_2|\sqrt{NND(A_1)}-\sqrt{NND(A_2)}|\nonumber \\
& +\frac{w_3}{2}\Big(\sqrt{\frac{\Gamma(P_{\alpha A_1},P_{\alpha A_2})}{log 2}}+\sqrt{\frac{\Gamma(P_{\alpha A_1^c},P_{\alpha A_2^c})}{log 2}}\Big)
\end{eqnarray}

\emph{Kernel-based distance.--} A recently proposed distance is based on diffusion kernels~\cite{Hammond2013}. This method estimates the differences between diffusion patterns of two networks undergoing a continuous node-thermal diffusion. A set of distances at different scales $t$ can be obtained by means of the Laplacian exponential kernels $e^{-t\mathcal{L}_{\{A_1,A_2\}}}$. The kernel-based distance is obtained by:
\begin{equation}
D_k = \|\exp(-t\mathcal{L}_1)-\exp(-t\mathcal{L}_2) \|_F
\end{equation}
where $\mathcal{L}_k$ denotes the graph Laplacian of  network $k$.

To assess the performances of these functions to capture network's differences, we consider a network $A$ and a set of perturbed networks $\{A_p\}$ generated with an incremental rewiring probability $p$ of original network $A$. We evaluate $f$'s by computing the differences between perturbed versions $\{A_p\}$ and its original configuration $A$. For low values of $p$ networks are very similar. Network differences are expected to increase with $p$.

\paragraph*{Benchmark tests.--}
We build binary Barabasi-Albert (BA) and Strogatz-Watts (SW) \cite{Newman2010} models with $L$ links and $N=100$. For SW model, the number of initial neighbors is $K=4$ for a $L=N*K$ edges and $\langle k\rangle=2K$. For each model we recreate a continuous perturbation process by reshuffling their links with and incremental rewiring probability step $p=0.001$. This allows us to create a set of $||\{A_p\}||=1000$ connected networks, each of them with $L*p$ rewired links. 

\begin{figure*}[ht]
\centering
\resizebox{0.95\textwidth}{!}{\includegraphics{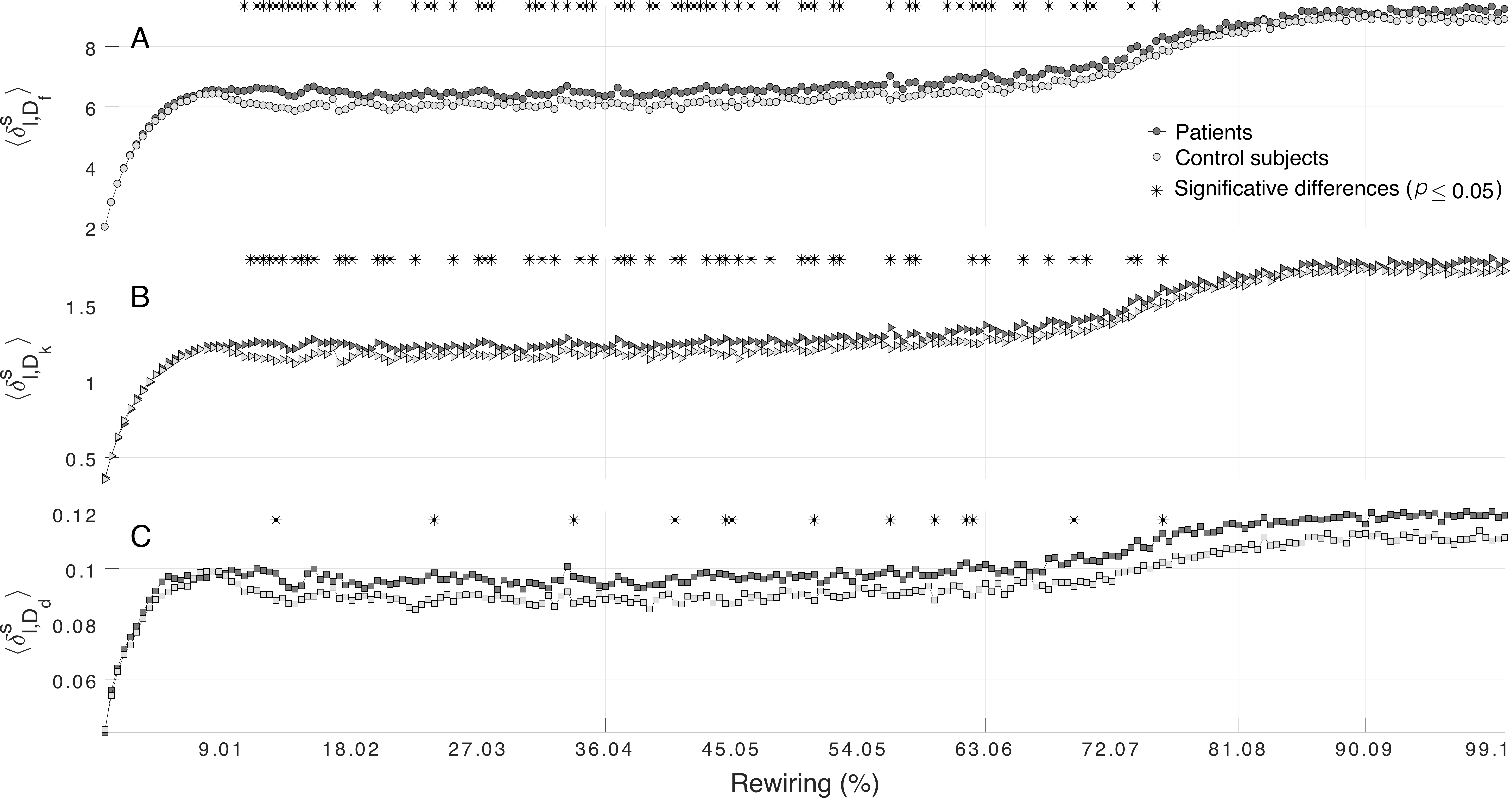}}
\caption{Mean distances profiles $\langle\delta_{l,f}^s\rangle$ for $P$ (black) and $C$ (white) are plotted as a function of the rewired $l$ links (Rewiring \%). The existence of group statistical differences in each rewiring step are highlighted at the top of each panel by the black stars.  \textbf{A}: Euclidean distance yields a  discriminability of 32.89 \%.  \textbf{B} the kernel-based distance yields a discriminability of 26.58\%. in \textbf{C}: for $f=D_d$ the discriminability performances are poor in comparison with the others metrics.}
\label{fig:02}
\end{figure*}

Let $\delta_{p,f}$ be the network-distance vector that contains all differences between perturbed networks $\{A_p\}$ and A measured for a given metric $f$. We compute the averaged profiles $\langle\delta_{p,f}\rangle$ as well as the 5$^{th}$-95$^{th}$ percentiles (Fig. \ref{fig:01}). All the averaged profiles display monotonically increasing curves that reach out certain saturation around $p=10^{-1}$.  Results suggest that all the measures (including the Euclidean distance) are sensitive to small structural changes (10\% of reshuffled links), and reflect well the structural perturbations. Beyond this threshold ($p> 10^{-1}$), however, all functions cannot distinguish between a graph $A$ and its perturbed version $\{A_p\}$. 

\paragraph*{Assessment of performances.--} To assess the metrics' performances we quantify the ``discriminability'' and the ``runtime''. Discriminability assesses whether a given function $f$ is sensitive at certain perturbation $p$, and whether it is suitable for distinguish two different group of networks at a given $p$. Discriminability is defined as the percentage of times a function $f$ distinguishes the differences of each group of networks at certain perturbation level. The more times $f$ distinguishes two different datasets, the better the $f$ discriminability is. In addition, runtime simply measures the $f$ execution time. The faster a given function $f$ estimates the differences, the better the corresponding metric is. For the sake of applicability we tested the performance of different $f$'s in real networks.

\paragraph*{Real networks.--} In this work, we use a recently published brain connectivity dataset, which includes functional connectivity matrices estimated from magnetoencephalographic (MEG) signals recorded from 23 Alzheimer patients ($P$) and a set of controls subjects ($C$) during a condition of resting-state with eyes-closed~\cite{Guillon2017}. Alzheimer disease is caracterised by anatomical brain deteriorations, which are reflected in an abnormal brain connectivity. MEG activity was reconstructed on the cortical surface by using a source imaging technique~\cite{Guillon2017}.  Connectivity matrices were obtained from $N=148$ regions of interest by means of the spectral coherence between activities in the band of  11-13~Hz. We specifically focused on this frequency band, which is particularly activated during resting activity with closed eyes, and it reflects the main functional connectivity changes accompanying the disease~\cite{Stam2002}. All the recording parameters and pre-processing details of connectivity matrices are explained in Ref.~\cite{Guillon2017}.

Following the procedure of Ref.~\cite{DeVicoFallani2017}, we thresholded each connectivity matrix by recovering its minimum spanning tree and then filling the network up with the strongest links until to reach a mean degree of three. This method is useful for optimizing the balance between the network efficiency and its rewiring cost~\cite{DeVicoFallani2017}. The resulting connectivity networks are binary adjacency matrices with $N=148$ nodes with $L=222$ links.

A direct comparison of connectivity matrices between the graphs of two groups $A \in \{P \lor C\}$ does not not allow to distinguish them. This result agrees with a previous studies that found group differences related to very local changes in connectivity~\cite{Stam2002, Guillon2017}. Authors in Ref.~\cite{Guillon2017} for instance, found that only 3\% and 4\% of the nodes accounts for the connectivity differences between groups, when different frequency bands are combined in the analysis. 

We propose an approach that allows to detect global network differences between those groups. For this, each connectivity graph $A$ is firstly perturbed by randomly choosing $l$ links $\forall\ l=1,2,\dots,L$ and reshuffling them such that the graph remains connected. We get thus a set of $||\{A_l\}||=222$ perturbed networks. We then compute the network differences between all pairs $(A,A_l)$. We finally repeat this procedure for 20 independent realizations. The distances profile $\delta_{l,f}^s$ results from the average of the network differences across realizations for a given subject $S$. The set of $||\{\delta_{l,f}^s\}||=23$ distances profiles per group (one for each subject) is used to compare the differences captured by $f$ when $l$ links are rewired. A function $f$ distinguishes two populations $\{\delta_{l,f}^s\}^P \wedge \{\delta_{l,f}^s\}^C$ at certain level $l$, if the group differences are statistically different at that perturbation level. Discriminability is defined as the hits percentage along all $L$ perturbations, i.e. the number of times the null hypothesis $H_o$ of no difference between the two groups is rejected. To assess significant differences, we used a non parametric permutation test allowing 500 permutations for each $l$ and we reject $H_o$ at $p \leq 0.05$ (corrected by a Bonferroni method). 

The mean distance profiles $\langle\delta_{l,f}^s\rangle$ for each $f$ are plotted in Fig. \ref{fig:02}. As in synthetic models, profiles show a monotonically increasing behaviour. At low rewiring percentages ($\leq 11\%$) there is no significant differences at group level. For small perturbation levels, functions $f$ can not distinguishes connectivity between groups. Something similar is observed when links perturbation are above $\approx 70\%$. On the other hand, $D_f$ appears as the one with the highest discriminability closely followed by $D_k$, while $D_d$ appears with lowest one. Results clearly suggest that Euclidean distance distinguishes better the two groups of networks considered here.

\begin{figure}[h]
\centering
\resizebox{0.9\columnwidth}{!}{\includegraphics{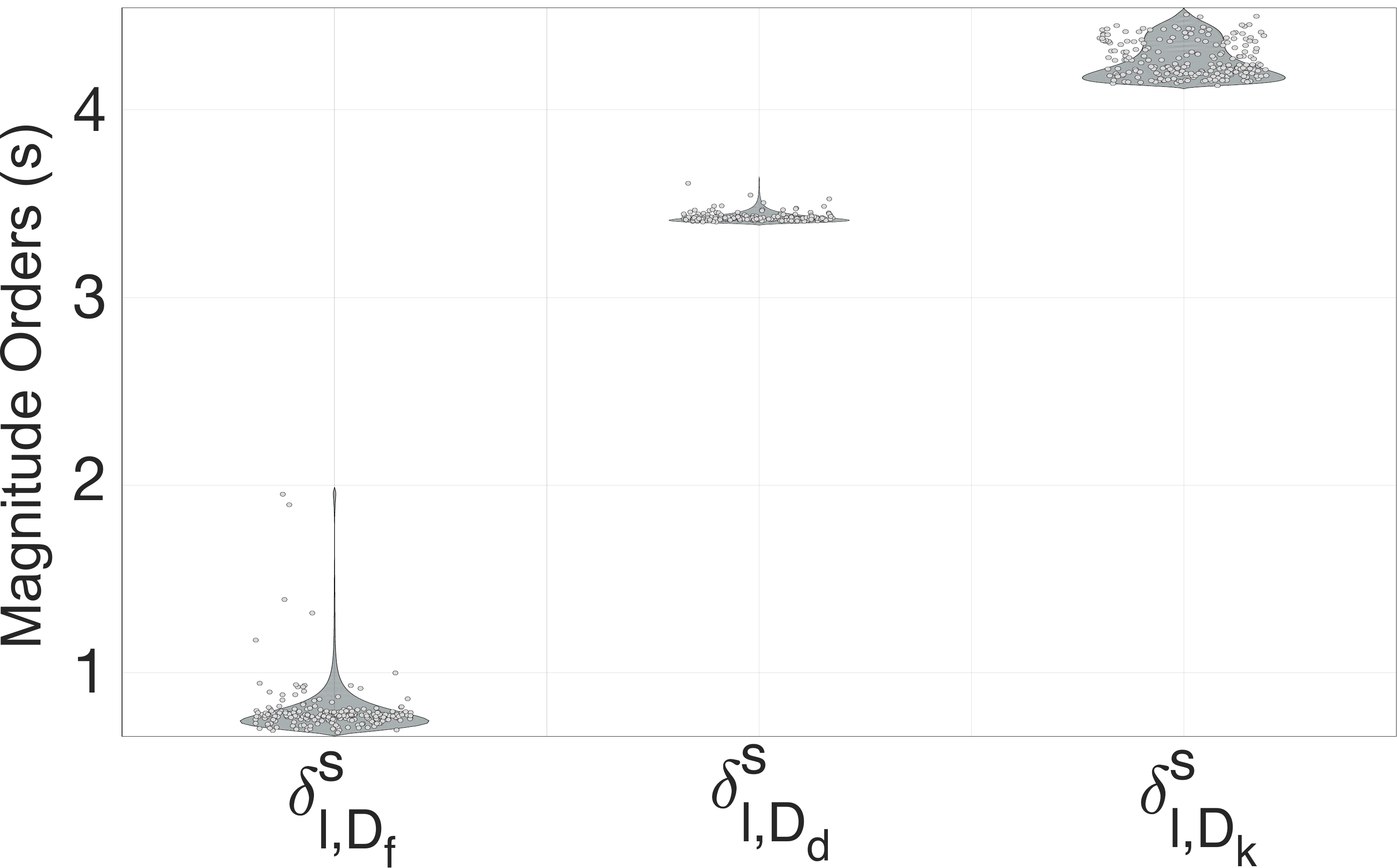}}
\caption{\label{fig:03} Relative orders of magnitude of execution times for three distance measures. Violin plots show the distributions of all values represented by the small circles. Although time differences between $D_d$ and $D_k$ is around one order, they become slower than $D_f$ execution time.}
\end{figure}

Finally, we assessed the execution time for computing a distances profile for each subject (we use MATLAB R2017a algorithms ran in an OS 10.12.6, with a 4GHz Intel dual core i7 processor, with 32GB of memory). Fig. \ref{fig:03} shows the relatives orders of magnitude in seconds that each metric takes to compute the networks differences. Average times obtained are: $t_f = 6.83\times10^{-5}$, $t_d = 2.68\times10^{-2}$, $t_k = 1.90\times10^{-1}$ for the Euclidean distance, the dissimilarity metric and the kernel-based method, respectively. The results clearly show Euclidean distance as the fastest method in comparison with the others two.  Clearly, $D_f$ is 3 (4) orders of magnitude faster than $D_d$ ($D_k$). 

Runtime finally determines which measure has the best performance when computing graphs distances. While the discriminability of $D_k$ is close to that of $D_f$, its runtime is four orders of magnitude slower than $D_f$ due to the fact that $D_k$ needs to search into several scales to find the highest difference. $D_d$ runtime is three orders of magnitude slower than $D_f$, because $D_d$ looks for many topological properties under several tuning parameters. In summary, the Euclidean distance emerges as the metrics with the best performances when computing graph differences. $D_f$ highlights for both: highest discriminability for distinguish groups of networks, as well as fastest computation, which is something really important when one manage large datasets.

\paragraph*{Concluding remarks.--}  Finding an accurate graph distance is a difficult task, and many metrics have been described without a framework to properly benchmark such proposals. Here we make a call of the simple Euclidean distance as the one with the best tradeoff between good and fast performances in contrast to more elaborated algorithms. In this work, we propose a method to detect global network differences with high efficiency and fast computation time. We also propose a simple framework to assess any metric's performance in terms discriminability and runtime. Results indicate that, for comparing binary networks of the same size, the Euclidean distance's discriminating capabilities outperform those of graph dissimilarity and diffusion kernel distance. More elaborated network models (e.g. multi-layer, weighted, signed or time-varying networks) might, however, need more elaborated tools to account for interdependencies of interacting units, and make their comparisons more robust.

\paragraph*{Acknowledgements.--} We are indebted to X. Navarro and M. Dovergine for their valuable comments.

\end{document}